\documentclass[fleqn,usenatbib,onecolumn]{mnras}

\usepackage{newtxtext,newtxmath}

\usepackage[T1]{fontenc}

\DeclareRobustCommand{\VAN}[3]{#2}
\let\VANthebibliography\thebibliography
\def\thebibliography{\DeclareRobustCommand{\VAN}[3]{##3}\VANthebibliography}

\usepackage{graphicx}	
\usepackage{amsmath}	
\usepackage[version=4]{mhchem}
\usepackage[left]{lineno}
\usepackage{bm}

\title[]{Potential Surface Ice Distribution on Close-in Terrestrial Exoplanets around M dwarfs}

\author[Ouyang \& Ding]{
Yueyun Ouyang,$^{1}$
Feng Ding$^{1}$\thanks{E-mail: fengding@pku.edu.cn}
\\
$^{1}$Laboratory for Climate and Ocean-Atmosphere Studies, Department of Atmospheric and Oceanic Sciences, School of Physics, Peking University, Beijing 100871, China.
}

\date{Accepted XXX. Received YYY; in original form ZZZ}

\pubyear{2024}

\begin{document}
\label{firstpage}
\pagerange{\pageref{firstpage}--\pageref{lastpage}}
\maketitle

\begin{abstract}
Previous studies suggested that surface ice could be distributed on close-in terrestrial exoplanets around M-dwarfs if heat redistribution on the planets is very inefficient. In general, orbital and atmospheric parameters play an important role in the climate on terrestrial planets, including the cold-trap region where the permanent surface water reservoir can potentially be distributed.  
Here, we develop a simple coupled land-atmosphere model to explore the potential surface ice distribution on close-in terrestrial planets with various orbital and atmospheric parameters, assuming that the planets are airless or have a thin \ce{N2} atmosphere. We find that the most significant factors in deciding the surface cold trap region are the spin-orbit ratio and obliquity. The incident stellar flux and the surface pressure play a limited role in the thin \ce{N2} simulations for incident flux smaller than Mercury's and surface pressure lower than 10$^4$ Pa. Our result illustrates the possible distribution of surface ice on arid terrestrial planets and can help to understand the climate of these exoplanets.
\end{abstract}




\section{Introduction}

The successful launch of the James Webb Space Telescope (JWST) opens a new era for exploring the climate of terrestrial planets around M dwarfs \citep{morley2017jwst,batalha2018jwst,krissansen2018trappist1e,lustig2019trappist1,gialluca2021jwst,wordsworth2022ARAA}. Close-in terrestrial planets, such as TRAPPIST-1 b and c, have become key targets for atmospheric detection with JWST due to their close proximity to parent stars, and preliminary observations suggest that both TRAPPIST-1b and c have thin atmospheres or are airless \citep{tr_b,tr_c}. 

Close-in M dwarf planets are subject to strong tidal dissipation, whose rotation states are likely to be in spin-orbit resonance \citep{goldreich1966spinorbit, heller2011tidal, barnes2017tidal}, making their insolation pattern similar to Mercury rather than Earth. Previous studies suggest that surface water ice is allowed to accumulate in the cold surface region (usually referred to as the surface cold trap) even for planets receiving stellar flux that is several times higher than Earth's \citep{leconte2013bistable,ding2020arid,ding2022arid}, when the atmosphere has low infrared (IR) opacity and therefore cannot redistribute heat efficiently \citep{wordsworth2015heat,koll2016}. Long-lived liquid water has also been proposed to possibly be present near the edge or at the bottom of the ice cap \citep{leconte2013bistable,lobo2023terminator}, raising an intriguing question regarding the regional habitability of such types of close-in exoplanets. 

In this paper, our aim is to explore how the potential surface ice distribution on close-in terrestrial exoplanets is impacted by orbital and atmospheric parameters with a simple coupled land-atmosphere model. We briefly introduce our land-atmosphere model and the setup of the model in Section~\ref{sec:method}. We then apply it to airless M dwarf planets in Section~\ref{subsec2} and to planets with thin \ce{N2} atmospheres in Section~\ref{subsec3}. We present our conclusion in Section~\ref{sec13}. 

\section{Methods} \label{sec:method}

\subsection{Coupled land-atmosphere model} \label{sec:land-atm}

We develop a simple coupled land-atmosphere model that calculates the evolution of atmospheric, surface, and subsurface temperatures over a full orbit with various orbital parameters, assuming that the atmosphere is transparent in IR with a low \ce{N2} content. We assume that the surface of the planet is made of rocks, with thermal conductivity $\kappa_T=2.9\hspace{0.5em}{\rm W~m^{-1}~K^{-1})}$ and thermal diffusivity $D=1.43\times10^{-6}\hspace{0.5em}{\rm m^2~s^{-1}}$ \citep[Chapter 7.4]{pierrehumbert2010ppc}. {We assume that the simulated terrestrial planet has the same mass as Earth with a gravity} value $g=9.8\hspace{0.5em}{\rm m~s^{-2}}$ and a uniform surface albedo $\alpha=0.2$. When there is no atmosphere at all, our model is very similar to the thermal model used to study the thermal phase curve of super-Mercury-type planets in \cite{selsis2013suermercury}. 

We first test our model with the airless scenario and select six input parameters: star type, average stellar flux ($F_{star}$), eccentricity (\textit{e}), spin-orbit ratio ($T_{ratio}=T_{orb}/ T_{rot}$), obliquity ($\gamma$), and internal heat flux ($F_{internal}$) {, where $T_{orb}$ and $T_{rot}$ are the orbit and rotation periods of the planet, respectively}. Specifically, we apply the luminosity and mass values for M1, M5, and M9 type stars and ensure that the incident stellar flux on the planets is the same as Mercury's or Earth's by varying the planet's semi-major axis (\textit{a}). In simulations with thin \ce{N2} atmospheres, we investigate two additional variables: surface pressure ($p_0$) and average surface wind speed (\textit{U}). 

To better compare and estimate the effects of each parameter on the surface temperature distributions, we choose a reference simulation: tidally locked planet around a M5 type star with Earth-like incident stellar flux, circular orbit with 1:1 spin-orbit ratio, uniform internal heat flux of 90 mW~m$^{-2}$, zero obliquity and precession angle. For the thin \ce{N2} atmosphere scenario, the reference surface pressure is $10^4$ Pa and the surface wind speed is 3 m~s$^{-1}$. For each numerical experiment, we change one parameter based on the reference simulation and calculate the temporal evolution of the surface temperature distribution. Note that some of the input parameters in our simulations are not independent. For example, for eccentric orbits, the spin-orbit ratio cannot be 1:1 {when the planet's spin becomes pseudo-synchronised and aligned with the orbit} \citep{correia2010exop.booktidal,markarov2012spinorbit,penev2024tidal}.

Then we estimate the maximum coverage of permanent surface ice based on surface cold trap distributions in our simulations with the criterion that the surface ice deposit could stay for 10~Gyr against hydrogen escape (Section~\ref{sec:timescale}). All input parameters used in our simulations are listed in Table~\ref{tab:input}.

\begin{table}
    \centering
    \begin{tabular}{cc}
    \hline
    Parameters     & Values\\
    \hline
    Stellar mass ($M/M_\odot$) &  0.079, {0.162}$^{\star,\dagger}$, 0.50 \\
    Luminosity ($L/L_\odot$) & 0.0003, {0.003}$^{\star,\dagger}$, 0.041\\
    Semi-major axis ($a$) &  0.39$\sqrt{L/L_\odot}$, {$\sqrt{L/L_\odot}$}$^{\star,\dagger}$ AU\\
    Eccentricity ($e$) & {0}$^{\star,\dagger}$,0.2\\
    Spin-orbit ratio ($T_{ratio}$) & {1:1}$^{\star,\dagger}$, 3:2, 2:1 \\
    Obliquity ($\gamma$) & {0$^\circ$}$^{\star,\dagger}$, 23$^\circ$, 45$^\circ$\\
    Internal heat flux ($F_{internal}$) & {90}$^{\star,\dagger}$, 1000 mW~m$^{-2}$\\
    \hline
    Surface pressure ($p_0$) & 0$^{\star}$, 100, 1000, {10$^4$}$^{\dagger}$ Pa\\ 
    Average surface wind speed ($U$) &  0$^{\star}$, {3}$^{\dagger}$, 6 m~s$^{-1}$\\
    \hline
    \end{tabular}
    \caption{Parameters used in our simulations. Stars mark the values for the reference simulation on airless planets, and daggers for planets with thin \ce{N2} atmospheres.}
    \label{tab:input}
\end{table}

\subsection{Incident stellar flux}
We assume that the simulated planets revolve around their parent stars with Keplerian orbits. The incident stellar flux at any point on the surface of the planets can be calculated by using Kepler's law and spherical geometry \citep[Chapter 7.5]{pierrehumbert2010ppc}. For a planet orbiting around a star with the mass of \textit{M} and the semi-major and semi-minor axis of \textit{a}, \textit{b}, its period $T_{orb}$ and the angular momentum $J$ of the revolution are
\begin{equation}
    T_{orb}=\sqrt{\frac{4\pi^2a^3}{GM}}, J=\frac{2\pi ab}{T_{orb}}.
\end{equation}
Because the spin-orbit ratio $T_{ratio}$ is given as input parameters in our simulation, the period and angular velocity of rotation $T_{rot}, \omega$ can be calculated as $T_{rot}=2\pi/\omega=T_{orb}/T_{ratio}$. Suppose that the planet is at the northern hemisphere solstice when $t=0$, and $\kappa$ is the season angle that describes its position relative to the initial one. The evolution of the season angle $\kappa(t)$ can be derived by the conservation of angular momentum as
\begin{equation}
    \frac{d\kappa}{dt}=\frac{J}{r^2}=\frac{J}{[a(1-e)]^2}(1+e\cos{\kappa})^2.
\end{equation}
Then, the longitude and latitude of the substellar point on the surface of the simulated planet ($\lambda_0, \phi_0$) can be calculated by
\begin{equation}
    \lambda_0(t)=\lambda_0(t=0)-\int_{0}^{t}\left(\omega-\frac{d\kappa}{dt}\right)dt', \phi_0(t)=\arcsin[{\cos{(\kappa-\theta)}\sin{\gamma}]}
\end{equation}
where $\theta$, $\gamma$ is the precession angle and obliquity. The zenith angle $\zeta$ at any point on the surface of the planet ($\lambda, \phi$) can be calculated by
\begin{equation}
    \cos{\zeta(t)}=\cos{\phi}\cos{\phi_0(t)\cos{[\lambda-\lambda_0(t)]}+\sin{\phi}\sin{\phi_0(t)}}.
\end{equation}
For $\cos{\zeta(t)<0}$, the star is below the horizon and no incident stellar flux can reach the surface. Let the star's luminosity be \textit{L}. The incident stellar flux at any point on the surface of the planet is
\begin{equation}
    S(t)=\frac{L}{4\pi r^2}\cos{\zeta(t)}.
\end{equation}
In our simulations, the integration time step is $\Delta t=T_{rot}/200$. {The horizontal grid resolution is $18^{\circ}\times18^{\circ}$}. We integrate our model for 50 planetary years to ensure that the simulated climate reaches equilibrium.

\subsection{Heat conduction in subsurface layer} \label{sec:energy equation}

As the vertical temperature gradients in the subsurface layer are typically much greater than the horizontal one, only vertical heat conduction is taken into account in our simulations. The energy equation in the subsurface layer is 
\begin{equation}
    \partial_t (\rho_sc_{ps}T)=\partial_z(\kappa_T\partial_zT).
\end{equation}
Here $\rho_s$, $c_{ps}$ is the density and the specific heat capacity of the subsurface layer. The lower boundary condition is that the diffusive heat flux there is balanced by the internal heat flux $F_{internal}$. 

For the boundary condition at the surface in the airless scenario, the diffusive heat flux at the surface is balanced by the net heating by insolation and thermal emission
\begin{equation} \label{eq:energy_airless}
    \kappa_T\partial_zT|_{z=0}=(1-\alpha)S(t)-F(T_{s}).
\end{equation}
The infrared emissivity of the surface is assumed to be unity and $F(T_s)=\sigma T_s^4$ where $T_s$ is the surface temperature. We set 35 layers in the subsurface layer and the depth of each layer \textit{i} is discretized as
\begin{equation}
    z_i=Z\frac{e^{i/5}-1}{e^3-1}, i=1,2,...,35
\end{equation}
where $Z=\sqrt{\kappa_T \Delta t/\rho c_{ps}}$, which is the characteristic depth to which temperature fluctuation can penetrate within one time step. The maximum depth $z_{35}$ in our simulations with various orbital parameters is within 10 metres.

\subsection{Reproducing the surface temperature evolution of the Moon and Mercury} \label{sec:moonmercury}

\begin{figure}%
\centering
\includegraphics[width=0.95\textwidth]{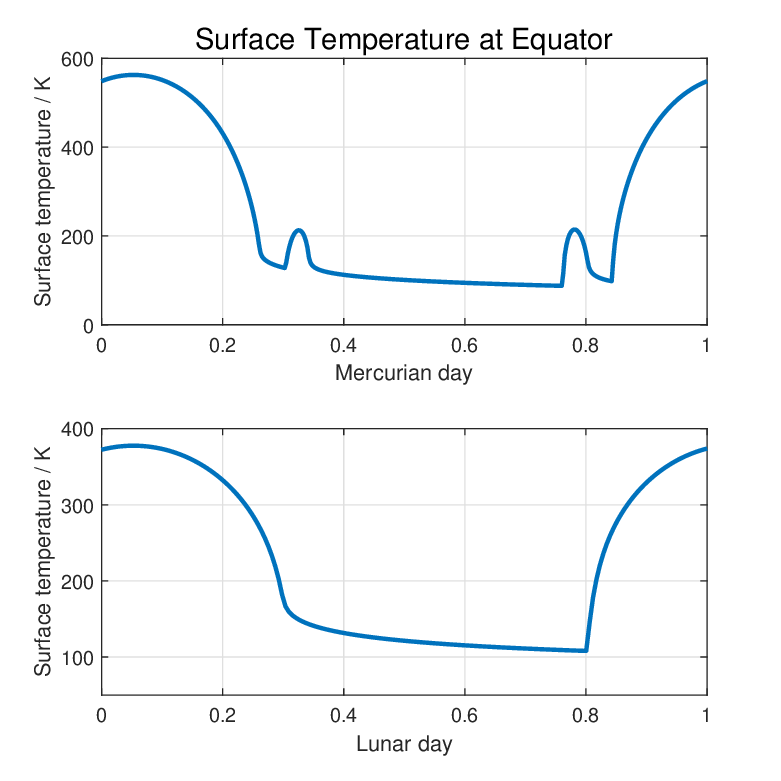}
\caption{Simulated surface temperature at the equator of Mercury (top) and Moon (bottom) as a function of local time by our simple land-atmosphere model. The Mercury results are for 90$^\circ$E longitude, where secondary sunrise and sunset happen due to the faster orbital angular velocity than its spin rate at perihelion.}\label{fig:moon}
\end{figure}

To validate our model in the airless scenario, we use our model to simulate the surface temperature evolution on the Moon and Mercury. For the soil heat conduction equation, the lunar regolith thermal parameters ($\kappa_T=0.01~{\rm W~m^{-1}~K^{-1}}, D=10^{-8}~{\rm m^2~s^{-1}}$) are used for the Moon and Mercury \citep[Chapter 7.4]{pierrehumbert2010ppc}. The surface albedo is set to be globally uniform, with 0.136 for the Moon and 0.088 for Mercury \footnote{\url{https://nssdc.gsfc.nasa.gov/planetary/planetfact.html}}, and the horizontal grid resolution is $18^{\circ}\times18^{\circ}$. 

The simulated surface temperature evolution at the equators of the Moon and Mercury (Fig.~\ref{fig:moon}) is very similar to previous work with similar thermal properties of the regolith \citep{vasavada1999mercury}. The Moon's surface temperature at the equator varies from $\sim 390$~K at noon to $\sim 100$~K at night. On Mercury it varies from $\sim 570$~K at noon to $\sim 100$~K at night. The secondary sunrise and sunset at specific longitudes on Mercury, due to the faster orbital angular velocity than its spin rate at perihelion, are also correctly captured in our simulation. The surface temperature variation in our simulations may be slightly different from the observed values, because the thermophysical properties of the regolith are depth- and temperature-dependent \citep{vasavada1999mercury,hayne2017moonregolith}. 

\subsection{Estimation of surface cold-trap distribution} \label{sec:timescale}
We use a critical temperature to estimate the permanent surface cold-trap distribution assuming that an ice layer with thickness $d_{ice}$ of 1~km could stay for 10 Gyr against hydrogen escape.  
For the airless scenario, we simply assume that all the water molecules sublimated from the ice surface can leave the planet immediately by
photodissociation into hydrogen molecules and subsequent hydrogen escape. Thus, the loss rate of the surface ice sheet depends only on the sublimation rate
\begin{equation}
    F_{eva}=\alpha\rho_sv_s(2\pi)^{-\frac{1}{2}} \left(1-\frac{p_0}{p_v} \right)
\end{equation}
where sticking coefficient $\alpha=1$, $\rho_s=p_v/R_{\ce{H2O}}T$ is the density and $v_s=(kT/m_{\ce{H2O}})^{1/2}$ is the molecular speed of water vapour in equilibrium with the surface ice, the background atmosphere pressure $p_0=0$ in the airless scenario and the water vapour pressure is equivalent to the saturation vapor pressure $p_v=p_s(T_s)$ \citep{ingersoll1985}.We use the empirical formula \citep{huang2018} to estimate the saturation vapor pressure in phase equilibrium with the surface ice with temperature $t$ in Celsius
\begin{equation}
    p_{s}(t)=\frac{\exp(34.494-\frac{6545.8}{t+278})}{(t+868)^2}\qquad(t = T_s - 273.15 \leq 0^\circ{\rm C}).
\end{equation}

In this way, the loss rate of the surface ice sheet can be expressed as a function of surface temperature. For a conservative estimate, we assume that the ice sheet can last for more than $t_{ice}=$ 20 Gyr, attaining the temperature threshold in the airless scenario with $T_0\approx140$ K by $F_{eva}(T_0)\leq\rho_{ice}d_{ice}/t_{ice}$.

For the case with a thin \ce{N2} atmosphere, we use the diffusion-limited escape flux of hydrogen to estimate the loss rate of surface ice sheet \citep{leconte2013bistable,catling2017evolution}:
\begin{equation}
    F_{esc}=f_{str}(\ce{H2})b_{ia}\frac{(m_a-m_{\ce{H2}})g}{kT_{str}}.
\end{equation}
Here $f_{str}$ is the mixing ratio of \ce{H2} and $T_{str}$ is the temperature in the upper atmosphere, $b_{ia}=1.9\times10^{21}(T/300K)^{0.75}$ $\rm m^{-1}s^{-1}$ is the binary diffusion parameter for \ce{H2} in the \ce{N2} atmosphere, $m_a, m_{\ce{H2}}$ is the molecular mass of \ce{N2} and \ce{H2}. The mixing ratio of hydrogen in the atmosphere can be estimated approximately as the saturation value of surface water vapour, because the potential ice layer cannot be stably trapped on the surface with an effective tropopause cold trap \citep{ding2020arid}. $T_{str}$ is estimated by the skin temperature of the planet. 

We also assume efficient horizontal mixing of water vapour in the atmosphere. Therefore, the loss rate of the surface ice sheet is relevant to the area ratio of the surface ice sheet. Then the lifetime of surface ice sheet can be calculated as
\begin{equation}
    t_{ice}=\frac{\rho_{ice}d_{ice}}{m_{\ce{H2O}}F_{esc}} \frac{S_{ice}}{S_{esc}}
\end{equation}
where hydrogen escape is assumed to happen on a global scale with $S_{esc}=4\pi R_p^2$. The area ratio of surface ice $S_{ice}/S_{esc}$ can be determined by the cold trap region with a surface temperature lower than the critical temperature over a full orbit. Given that $t_{ice}\geq10$ Gyr, the critical temperature can be solved by the surface temperature distribution in simulations with various parameters.

\subsection{Energy conservation with a thin nitrogen atmosphere}

For an IR-transparent atmosphere, the atmosphere can only exchange energy with the underlying surface by the turbulent sensible heat flux \citep{pierrehumbert2011gl581g}, {and the atmosphere can still transport heat horizontally by air flow.} Due to the long non-dynamical (e.g., radiative) timescale in the pure \ce{N2} atmosphere {relative to the short dynamical adjustment timescale}, the weak-temperature-gradient (WTG) approximation can be applied to the atmosphere. {Using the WTG approximation to implicitly solve for horizontal heat transport in IR-transparent atmospheres on synchronously rotating planets was first proposed in \citet{pierrehumbert2011gl581g}, and} verified by recent global climate modelling by \cite{wang2022puren2}. With the WTG approximation, the near-surface atmosphere has a horizontally uniform temperature $T_a$ and the energy conservation equation for the atmosphere is
\begin{equation}
    \frac{p_0}{g}c_{pa}\frac{dT_a}{dt}=a({T_s}-T_a).
\end{equation}
Here ${p_0}$ is the surface \ce{N2} pressure, $c_{pa}$ is the specific heat capacity of \ce{N2}, Parameter \textit{a} is the turbulent coupling coefficient calculated as $\rho_ac_{pa}C_dU$, where $\rho_a=p_0/R_{\ce{N2}}T_a$ is the surface air density, $U$ is the average wind speed, and $C_d$ is a dimensionless drag coefficient, with a typical value of 0.0015 over moderately rough surface \citep{Garratt1994}.

The sensible heat flux also changes the upper boundary condition of the heat conduction in the subsurface layer compared to that of the airless case
\begin{equation}
    \kappa_T\partial_zT|_{z=0}=(1-\alpha)S(t)-\sigma T_s^4-a({T_s}-T_a)\label{soil boundary 2}.
\end{equation}
Compared to the surface energy budget of typical terrestrial climates, the thermal emission flux of the atmosphere is ignored in Eq.~\ref{soil boundary 2} because the atmosphere is assumed to be transparent in IR. However, the existence of sublimated water vapour from the potential ice surface may violate this assumption. 
To evaluate the greenhouse effect contributed by water vapour, a line-by-line radiation code (PyRADS-shortwave) is used \footnote{\url{https://github.com/danielkoll/PyRADS-shortwave}}, which has been used to study the greenhouse effect of water vapour on Earth and other exoplanets \citep{koll2018,Koll_2019}. Assuming a background surface pressure of $10^4$ Pa, a vertical uniform temperature profile of 200~K and water vapour profile with the saturation value at the surface, we find that the downward infrared flux reaching the surface is less than 6~W~m$^{-2}$, which is one order of magnitude lower than the maximum sensible heat flux in our reference simulation. This comparison confirms that our assumption with zero IR opacity can be treated as a first-order approximation, although the simulated surface temperature by Eq.~\ref{soil boundary 2} may be slightly underestimated if the greenhouse effect of sublimated water vapour is taken into account. 

{We also use PyRADS-shortwave to evaluate the shortwave radiative effect of the $10^4$ Pa \ce{N2} atmosphere with low \ce{H2O} concentrations, because \ce{H2O} has plenty of absorption lines in the near-infrared and can potentially decrease the planetary albedo while the Rayleigh scattering by \ce{N2} can slightly increase the planetary albedo. For a late M-dwarf with an effective temperature of 2500~K, 1.2\% of incoming stellar flux is absorbed in the atmosphere by \ce{H2O} and the planetary albedo is reduced by 0.3\% relative to the surface albedo. For an early M-dwarf with an effective temperature of 3700~K, only 0.65\% of incoming stellar flux is absorbed in the atmosphere by \ce{H2O} and the planetary albedo is reduced by 0.05\%. The radiative calculations suggest that the stellar energy distribution has a trivial effect on the energy budget of the terrestrial planets with thin atmospheres, and validate our assumption that ignores the shortwave radiative effect of the thin atmospheres in this work.}

\section{Airless Scenario}\label{subsec2}

\begin{figure}%
\centering
\includegraphics[width=0.95\textwidth]{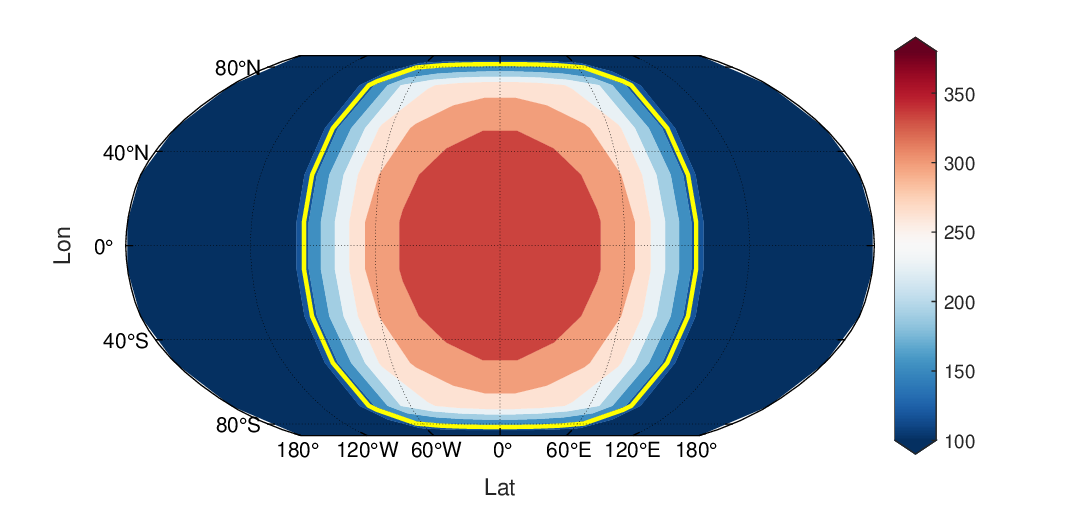}
\caption{Surface temperature distribution in the reference simulation on airless planets (unit: Kelvin). The yellow contour marks the critical temperature of $\sim$140 K permitting the existence of permanent surface ice that spans the entire nightside.}\label{ref_state1}
\end{figure}

For airless planets, the surface can be warmed by either the incident stellar flux from above or the diffusive heat flux from below and cooled by thermal emission to space (Eq.~\ref{eq:energy_airless}). 
The surface temperature distribution in the reference simulation on airless planets is shown in Fig. \ref{ref_state1}. As the planet is in the synchronous rotation state with zero obliquity and eccentricity, the incident stellar flux stays constant at any point on the surface, resulting in a time-invariant equilibrium state. The hottest place coincides with the substellar point, reaching approximately 360 K as a result of intense incoming shortwave radiation. The coldest place occurs on the nightside, where the temperature drops below 100 K and is solely heated by the planet's internal heat flux. According to our estimate, the critical temperature to allow the existence of surface ice for 10 Gyr is about 140 K, indicating that permanent surface ice could potentially span the entire nightside in the reference simulation. 

\begin{figure}%
\centering
\includegraphics[width=0.95\textwidth]{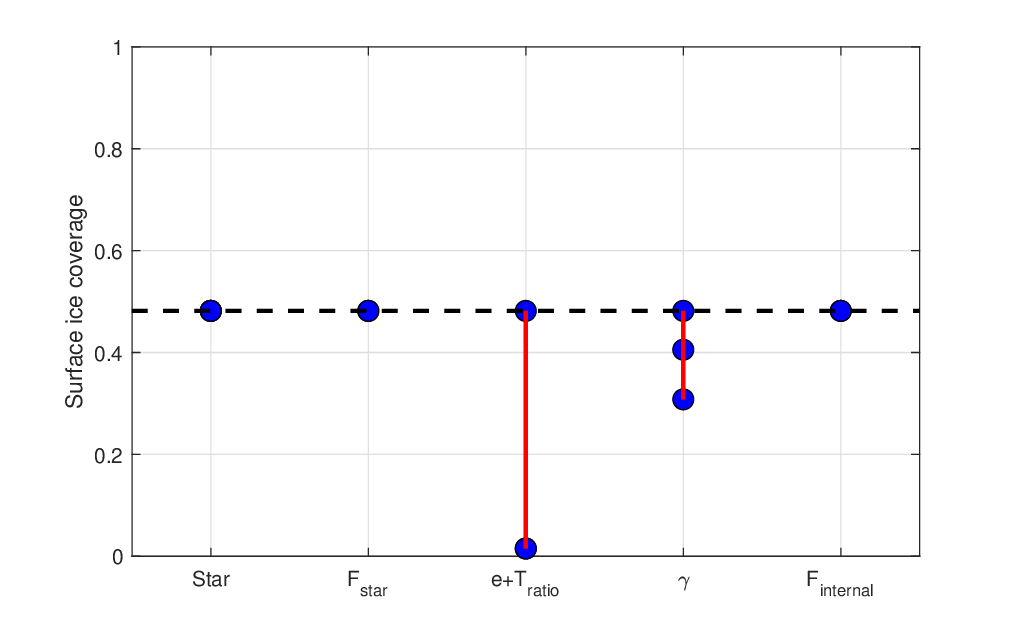}
\caption{{Sensitivity of the maximum coverage of permanent surface ice to the input parameters defined in Section~\ref{sec:land-atm}}. The dash horizontal line marks the maximum coverage of permanent surface ice in the reference simulation. The red vertical lines mark the range of ice coverage when the parameter listed in x-axis varies. {Note that the spin-orbit ratio and orbital eccentricity are not independent parameters and thus are taken into account together in the horizontal axis.}}\label{ice_rate1}
\end{figure}

Fig.~\ref{ice_rate1} shows the maximum coverage of permanent surface ice under different parameters listed in Table~\ref{tab:input}. In the reference state with a permanent nightside, the surface ice is allowed to spread throughout the night hemisphere and extend marginally to the dayside near the terminator where the insolation is very low. Therefore, the surface ice coverage is slightly higher than half. Among all the parameters we survey, only the influence of spin-orbit ratio, eccentricity, and obliquity is not negligible, while the star type, average stellar flux, and internal heat flux have a trivial effect. Non-1:1 spin-orbit ratio with non-zero eccentricities or non-zero obliquities can expose more surface to incident stellar radiation and warm the exposed surface. Note that the permanent surface ice distribution in our estimate requires that the surface temperature stays below 140~K over the course of a full orbit. 

For planets with a non-1:1 spin-orbit ratio, the substellar point can move along the zonal direction compared to the synchronous rotation state, which forces the permanent surface cold trap to retreat to the polar region. This is similar to Mercury's state with the presence of water ice near the north pole despite the fact that this planet receives an averaged stellar flux seven times as high as the Earth \citep{lawrence2013mercury}.   

Non-zero obliquity also plays an important role in the evolution of the substellar point on the planet, leading to the motion of the substellar point along the meridional direction and the seasonal variation in both the northern and southern hemispheres, similar to the seasonal cycles on Earth or Mars.  When obliquity increases, more surface area near the poles will be exposed to the incident stellar flux in the summer season even with the 1:1 spin-orbit ratio, causing the permanent surface cold trap to retreat to mid and low latitudes on the nightside.

The incident stellar flux at the substellar point and the internal heat flux can change the surface temperature distribution, as energy inputs from the top and the bottom of the surface (Section~\ref{sec:energy equation}), respectively. However, the surface temperature of the nightside in our reference simulation is only relevant to the internal heat flux, which is not sufficient to warm the nightside to the critical temperature. The surface temperature near the terminator is also altered marginally by these two parameters. Thus, the incident stellar flux and the internal heat flux have a very trivial impact on the distribution of surface cold traps in our simulations. The star type has no effect on the surface temperature distribution in Fig.~\ref{ice_rate1}, because the incident stellar flux, spin-orbit ratio, and surface albedo of the planet are fixed as in our reference simulation, although the actual orbital period and rotation period can change. 

\section{Scenario with thin nitrogen atmospheres}\label{subsec3}

\begin{figure}%
\centering
\includegraphics[width=0.95\textwidth]{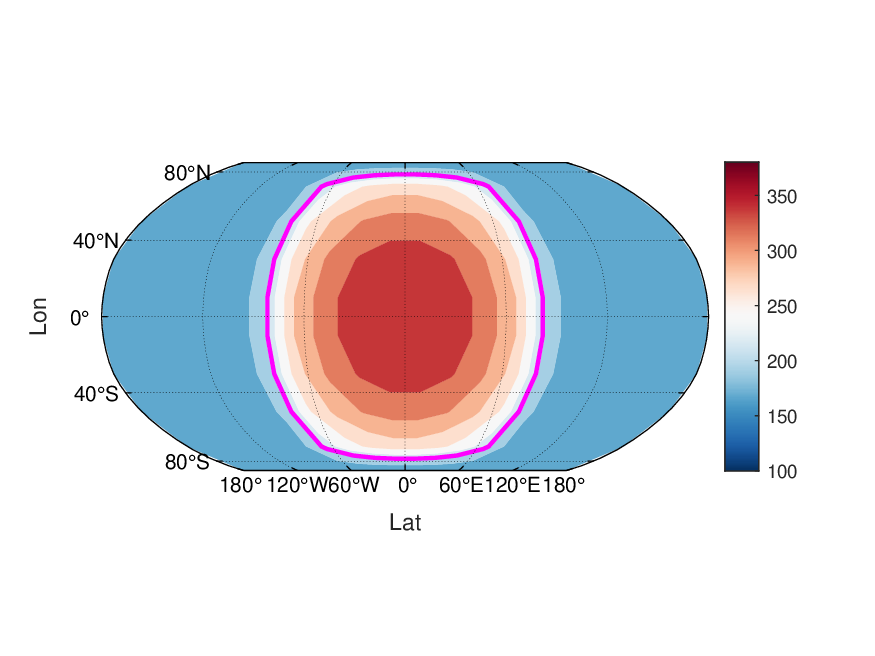}
\vspace{-60pt}
\caption{Surface temperature distribution in the reference simulation on planets with a thin \ce{N2} atmosphere (unit: Kelvin). The pink contour marks the critical temperature of $\sim$220 K permitting the existence of permanent surface ice.}\label{ref_state2}
\end{figure}

For planets with a thin \ce{N2} atmosphere, the surface temperature is subject to sensible heat transfer between the surface and the atmosphere, in addition to incident stellar flux, diffusive heat flux, and thermal emission flux in the airless scenario (Eq.~\ref{soil boundary 2}). The surface temperature distribution in the reference simulation with 10$^4$ Pa \ce{N2} and an average surface wind speed of 3~m~s$^{-1}$ is shown in Fig.~\ref{ref_state2}. Compared to the airless case in Fig.~\ref{ref_state1}, the highest temperature at the substellar point decreases, while the surface temperature on the nightside increases. This more uniform surface temperature distribution results from sensible heat transfer between the surface and the atmosphere, which cools the surface region hotter than the atmosphere and warms the surface region colder than the atmosphere. This heat redistribution mechanism is different from that dominated by radiative transfer in the thermal IR described in \cite{koll2016}.
Furthermore, the critical temperature to allow the existence of surface ice for 10~Gyr, indicated by the pink contour in Fig.~\ref{ref_state2} , notably increases to 220~K in our reference simulation, because nitrogen atmospheres effectively slow hydrogen escape and thus reduce the loss rate of the surface ice sheet. These combined factors drive the expansion of surface ice coverage to  $\sim$60\% under Earth-like insolation, further into the dayside hemisphere compared to the airless case.

\begin{figure}%
\centering
\includegraphics[width=0.95\textwidth]{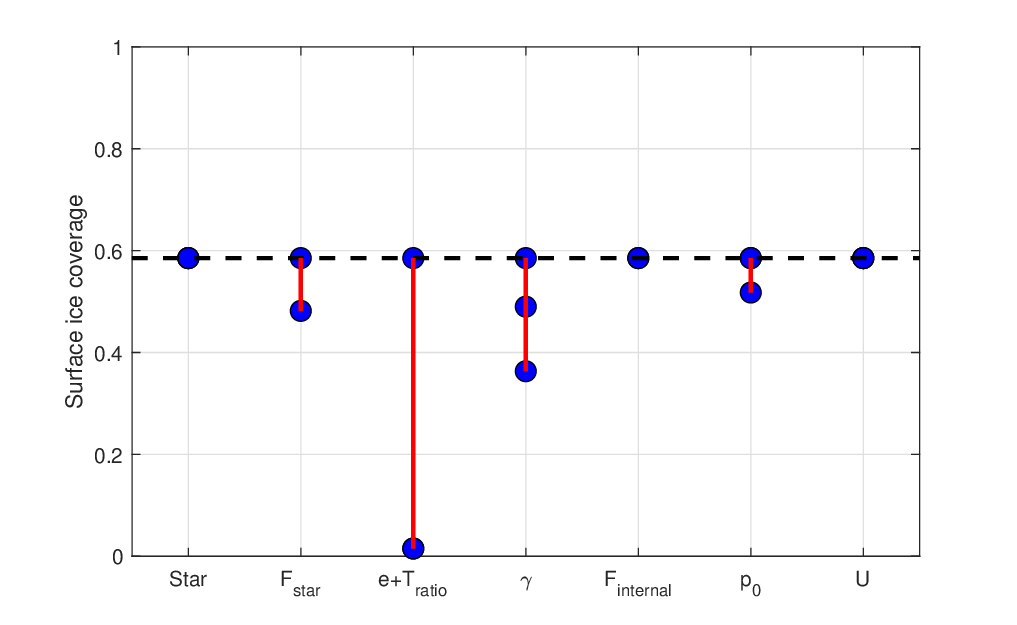}
\caption{Same as Fig.\ref{ice_rate1}, but for planets with thin \ce{N2} atmospheres. The surface pressure of \ce{N2} ($p_0$) and average surface wind speed ($U$) are added in the horizontal axis.}\label{ice_rate2}
\end{figure}

Fig.~\ref{ice_rate2} shows the maximum coverage of permanent surface ice for planets with thin \ce{N2} atmospheres. Similar to the airless case, the surface ice coverage reaches its maximum in the reference simulation with 1:1 spin-orbit ratio and zero obliquity. The spin-orbit ratio, obliquity, and eccentricity play a significant role in the surface temperature distributions. 

Unlike airless simulations, incident stellar flux can affect surface cold trap coverage in a minor way. When the planet receives incident flux as high as Mercury, the intense insolation warms the terminator region significantly while still keeping the nightside cold. It causes the surface cold trap region to retreat to the nightside, similar to the airless case. The effects of atmospheric parameters also appear to be relatively minor, although both surface wind speed and surface air pressure can affect sensible heat transfer between the surface and the atmosphere. For the typical values that we explore in the simulations, their impact is smaller than the orbital parameters we discussed above, even when the surface pressure is varied by two orders of magnitude.

\section{Conclusion and implication for planet survey}\label{sec13}

We investigate the potential surface ice distribution of close-in M dwarf planets using a simple coupled land-atmosphere model and discuss the effects of orbital and atmospheric parameters in airless and thin \ce{N2} atmosphere scenarios. In general, the 1:1 spin-orbit ratio and zero obliquity favour a global-scale distribution of surface cold traps with possible surface ice. For terrestrial planets that receive incident stellar flux less than that on Mercury and have a \ce{N2} atmosphere thinner than 10$^4$ Pa, stronger incident stellar fluxes and thinner atmospheres tend to decrease the surface coverage of cold traps, but in a limited way. 

Based on our simulation results, we filter the confirmed exoplanets around M dwarfs \footnote{\url{https://exoplanet.eu/}} and find several candidates that can potentially have a large-scale surface ice distribution on its nightside. We use the criterion with $r\le1.5r_\oplus$ and $m\le5m_\oplus$ to ensure that it is a terrestrial planet, eccentricity $e<0.05$ and the incident stellar flux between that of Mercury and Earth. The filtered result shows only several planets such as Proxima Centauri d and TRAPPIST-1 b-d can be hopeful candidates, and other close-in terrestrial planets either have unknown or large eccentricities, or are too far away from the Earth to be characterised. 

In this work, we only estimate the potential surface ice coverage by computing the surface cold trap distributions. Close-in planets may be water poor if they formed during the prolonged pre-main sequence phase of their parent stars \citep{ramirez2014,luger2015,tian2015}. Without a water supply for the nightside surface, the area can still be completely dry. Recent work discussed the possibility of secondary atmosphere buildup on close-in terrestrial planets by volcanic outgassing from the hydrated interior \citep{kite2020secondary}. Other outgassed molecules such as \ce{CO2}, \ce{CO} and \ce{CH4} can increase the IR opacity of the atmosphere and thus lower the chance of surface ice formation. Mineral dust is another complicating factor. Dust particles can be lifted to the atmosphere by dust activity on the dayside and transported to the nightside by large-scale circulation, where dust can increase the IR opacity and warm the nightside surface similar to greenhouse gases \citep{boutle2020dust}. However, recent GCM simulations suggested that the convective behaviour in the atmosphere on tidally locked planets, which is crucial to trigger dust activity, still requires a substantial background IR opacity \citep{wang2022puren2}. {At last, it should be noted that other mechanisms have been proposed to increase the water content on terrestrial planets around M dwarfs against atmospheric escapes, e.g., migration of planets in the proto-planetary disks \citep{alibert2017waterrich} and oxidation of atmospheric hydrogen by rocky materials from incoming planetesimals and from the magma ocean \citep{kimura2022water}.}

{Another simplification we made in this work is that the incident radiation from the host star reaches the top of the atmosphere on the terrestrial planet as plane-parallel rays, which works well for terrestrial planets in the solar system. However, for close-in terrestrial planets around their host stars, the finite angular size of stars may make more than 50\% of the planet receive the stellar radiation. This hyper-illumination effect is important for the climate dynamics of lava planets that receive intense stellar radiation because of their close proximity to the host stars \citep{nguyen2020k2141b,kang2023TPWlava}. Recently, \cite{Carter2024illumination} showed that even for a temperate terrestrial planet such as TRAPPIST-1 d, the illumination can reach 51.2\% and is still slightly higher than one-half. The hyper-illumination effect may make the surface ice distribution on close-in terrestrial planets around M dwarfs more sensitive to the orbital obliquity than discussed in our present work. } 
These factors discussed above should be taken into account in future studies to better understand the climate on close-in terrestrial planets.


\section*{Acknowledgements}

The authors thank the referee for thoughtful comments that improved the manuscript. The authors also thank Daniel Koll and Jun Yang for helpful discussions. The authors acknowledge funding support from the Fundamental Research Funds for the Central Universities (Peking University). 

\section*{Data Availability}

Data generated by this study including short movies on the surface temperature evolution on the Moon and Mercury are available at \url{https://doi.org/10.5281/zenodo.10682867}. The source code for our simplified land-atmosphere model is available on Github at \url{https://github.com/OuyangYueyun/Surface-ice}.



\bibliographystyle{mnras}
\bibliography{refs}{}








\bsp	
\label{lastpage}
\end{document}